\documentclass[aps,12pt,showpacs]{revtex4}
\usepackage{cmap}
\usepackage{graphicx}
\usepackage{color}
\usepackage{amsmath, amsfonts}

\definecolor{lightlightblue}{rgb}{.85,1,1}
\def\G{\mathcal{G}}
\def\ra{\rightarrow}

\def\be{\begin{equation}}
\def\ee{\end{equation}}

\def\bea{\begin{eqnarray}}
\def\eea{\end{eqnarray}}
\renewcommand\L{\mathcal{L}}
\def\G{\mathcal{G}}
\def\A{\mathcal{A}}

\def\O{\mathcal{O}}

\newcommand{\rf}[1]{(\ref{#1})}

\usepackage{graphicx}
\usepackage{epstopdf}
\begin{document}

\title{Non-contractible loops in the dense $O(n)$ loop model on the cylinder}

\author{F.C. Alcaraz$^1$, J.G. Brankov$^{2,3}$, V.B. Priezzhev$^2$, V. Rittenberg$^4$, A.M. Rogozhnikov$^5$ }
\affiliation{
$^1$Instituto de F$\acute{i}$sica de S$\tilde{a}$o Carlos, Universidade de S$\tilde{a}$o Paulo,
Casixa Postal 369, 13560-590, S$\tilde{a}$o Carlos, SP, Brasil \\
$^2$Bogoliubov Laboratory of Theoretical Physics, JINR, 141980 Dubna, Russia \\
$^3$Institute of Mechanics, Bulgarian Academy of Sciences, 1113 Sofia, Bulgaria \\
$^4$Physikalisches Institut, Universit$\ddot{a}$t Bonn, Nussallee 12, 53115 Bonn, Germany\\
$^5$National Research University « »Higher School of Economics, 101000 Moscow, Russia
}

\begin{abstract}
A lattice model of critical dense polymers $O(n)$ is considered for the finite cylinder geometry.
Due to the presence of non-contractible loops with a fixed fugacity $\xi$, the model at $n=0$ is
a generalization of the critical dense polymers solved by Pearce, Rasmussen and Villani.
We found the free energy for any height $N$ and circumference $L$ of the cylinder.
The density $\rho$ of non-contractible loops is obtained for $N \rightarrow \infty$ and large $L$.
The results are compared with those found for the anisotropic quantum chain with twisted boundary
conditions. Using the latter method we derived $\rho$ for any  $O(n)$ model and an arbitrary fugacity.
\end{abstract}

\maketitle


\section{Introduction}

\begin{figure}[!ht]
\includegraphics[width=60mm]{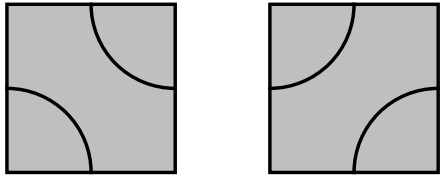}
\caption{\label{fig-cells} Elementary cells.}
\end{figure}

The dense $O(n)$ loop model \cite{dense} is defined by drawing two arcs in each
elementary cell of the square lattice: the two possible states of the cell are shown
in Fig. \ref{fig-cells}. The lines on the whole lattice with appropriate boundary conditions
form a system of closed loops with the Boltzmann weight $n$ ascribed to every loop.
\begin{figure}[!ht]
\includegraphics[width=80mm]{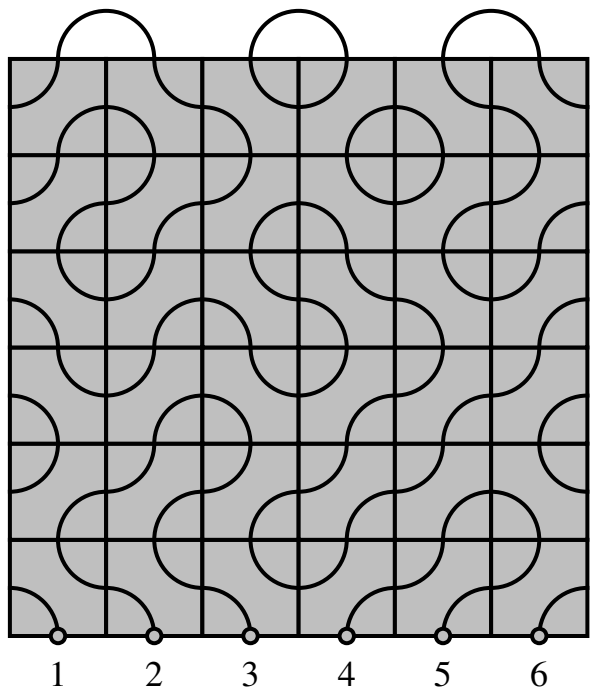}
\caption{\label{fig-loops} Loops on the horizontally periodic lattice. One noncontractible
loop joins the left and the right sides of figure.}
\end{figure}
Two particular cases, $n=1$ and $n=0$ are of special interest. At $n=1$, the model can be mapped to the bond percolation problem, the six vertex model and the $XXZ$ quantum chain \cite{Batch,Mitra}. At $n = 0$, the bulk loops disappear and the set of lines is converted into the model of critical
dense polymers \cite{Pearce}. This model is a recent representative of the more general two-dimensional polymer theory initiated by Saleur and Duplantier in the context of a conformal field theory \cite{Saleur-1,Dupl,Saleur-2}. The dense polymer
model is the first member $\mathcal{L}M(1,2)$ of the Yang-Baxter integrable series of
logarithmic minimal models. In the scaling limit, the central charge is $c=-2$ and the conformal weights yield the Kac formula for the infinitely extended Kac table \cite{Pearce}.

Pearce, Rasmussen and Villani \cite{PRVill} have solved the dense polymer model on a cylinder using the single-row transfer matrix and the inverse identity for the transfer matrix. They solved the
inverse identity for different boundary conditions, including the case in which noncontractible loops are allowed. The structure of the inversion identity dictates its solvability at fugacity 2 of the non-contractible loops. The fixed fugacity prevents the evaluation of the free energy as a function of fugacity, so the average number of loops on the cylinder remains unknown. At the same time, the statistics of noncontractible loops is important as it is related to the conformal properties of the model in the continuous limit. The standard tool for determination of the central charge $c$ of a conformal
field theory, corresponding to a given lattice model at the critical point, is the $1/L$ expansion of the free energy of an infinitely long cylinder of finite perimeter $L$ . The two leading terms of the expansion have the form
\begin{equation}
F=f_{\rm bulk}L-\frac{\pi c_{\rm eff}}{6 L},
\label{expansion}
\end{equation}
where $f_{\rm bulk}$ is the bulk density of the free energy per unit length, and $c_{\rm eff}$ is an effective central charge, which is a combination of the true central charge $c$ and a correction term depending on the boundary conditions \cite{Izmail}. It will be shown below, that the contribution of the non-contractible loops in $F$ is of the order of $1/L$. Therefore $c_{\rm eff}$ depends on the presence of such loops and their fugacity
in the case of two-dimensional lattice models, and on a twist parameter in the corresponding quantum chains.

In this paper, we solve the model of dense polymers on the cylinder by calculating the partition function of a spanning web model on a finite cylinder in the presence of cycles winding around the cylinder.
Our aim is to evaluate the grand partition function of the dense polymers model at arbitrary fugacity of the non-contractible loops and to find their density per unit height of the cylinder.

\begin{figure}[!ht]
\includegraphics[width=80mm]{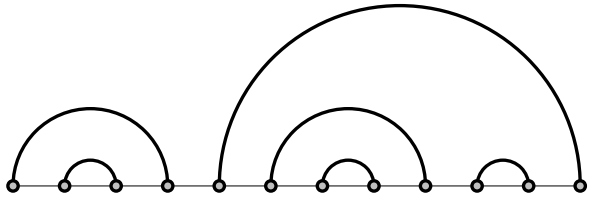}
\caption{\label{fig-links} A link configuration.}
\end{figure}

For an infinite cylinder of perimeter $L$, there is an alternative, albeit
simpler, method to compute the free energy and the density of the non-contractible
loops for any $O(n)$ model at arbitrary fugacity.
The states of the $O(n)$ model can be defined in terms of a connectivity condition for the points of intersection between the loops
and a horizontal line cutting the loops at these points. Two points are connected by a link if there exists a line between them via the half space above the cut. For instance, imposing periodic boundary conditions in horizontal direction on the lattice in Fig. \ref{fig-loops} and specifying the boundary conditions at the upper edge, we obtain three minimal links between points 1 and 6, 2 and 3, 4 and 5, which are points of intersection between loops and the bottom line of the lattice belonging to the upper half-plane. A typical configuration of links for a larger lattice is given in Fig. \ref{fig-links}.
The link representation allows definitions of the transfer matrix and the Hamiltonian as elements of
the periodic Temperley-Lieb algebra \cite{Levy,AAA}. The same algebra has a matrix representation that takes us
to the $XXZ$ spin-$1/2$ quantum chain with a twist depending on the fugacity.
The ground-state energy is known analytically \cite{Alcaraz} for large values of $L$ for any anisotropy related to the parameter $n$ of the $O(n)$ model. The Temperley-Lieb algebra has a quotient with a free parameter that can be identified with the fugacity of noncontractible loops, so we can compute their density
from the ground state of the Hamiltonian.  In the special case of $n =0$, one can use a Jordan-Wigner transformation and obtain the ground state energy for any finite $L$ explicitly.

The paper is organized as follows. In Section II, we present the
calculation of the density of noncontractible loops using the $XXZ$ quantum
chain. We start with this presentation since, unlike the lattice case, the
calculation is almost trivial. We also show that the probability distribution
of noncontractible loops is not Gaussian. In Section III, we consider the special case $n = 0$.
Due to the absence
of contractible loops, the $O(n)$ loop model can be mapped on the spanning
webs model. In the latter model, one can generalize the Kirchhoff theorem and
bring the calculation to that of determinants. The details of the proof are given in the Appendix.
The calculation of the partition function on a $L\times N$ torus
is presented in Section IV, and the result is given by Eq. (\ref{Z1}). The case of finite cylinders of height $N$ and perimeter $L$, with different boundary conditions on the top and bottom of the cylinder, is briefly mentioned in Section V. Finally, in Section VI, we consider the case of an infinite cylinder by
taking $N$ to infinity, and we compare the thus obtained results for the density of the noncontractible loops
to those of Section II.

\section{ The density of noncontractible loops obtained from the XXZ quantum chain}

  We remind the reader of a few facts about the {\it periodic} Temperley-Lieb algebra
(PTL) and some of its representations \cite{AAA}. This algebra provides the key to the calculation
of the density of non-contractible loops. The PTL has $L$ generators $e_i$
($i = 1,2,\ldots,L$) satisfying the relations
\begin{equation} \label{2.1}
  e_i^2 = x\, e_i,\quad  e_ie_{i\pm 1 }e_i =e_i,\quad e_ie_j = e_je_i\, (|i - j|>
1),\quad e_{i+L} =e_i,
\end{equation}
where $x$ is a parameter.

We are going to consider the $L$-even case only. The PTL is infinite-dimensional
therefore we take a  quotient which makes it finite dimensional:
\begin{equation} \label{2.2}
ABA =\alpha^2 A
\end{equation}
where
\begin{equation} \label{2.3}
A = \prod_{i = 1}^{L/2} e_{2i},
\quad B = \prod_{i = 0}^{L/2 - 1} e_{1+2i}.
\end{equation}
  We are interested in two representations of the PTL with the quotient \rf{2.2}. The
first one is the spin representation, in which
\begin{eqnarray} \label{2.4}
e_i = \sigma_i^+\sigma_{i+1}^-e^{i\phi/L} +
 \sigma_i^-\sigma_{i+1}^+e^{-i\phi/L} -\frac{\cos(\gamma)}{2} \sigma_i^z\sigma_
{i+1}^z \nonumber \\ +\frac{i}{2}\sin(\gamma) (\sigma_{i+1}^z-\sigma_i^z) +
\frac{\cos(\gamma)}{2},
\end{eqnarray}
where
\begin{equation} \label{2.5}
x= 2\cos(\gamma),\quad \alpha=2\cos(\phi/2),
\end{equation}
and $\sigma^{\pm,z}$ are the Pauli matrices \cite{Levy}, and $\phi$ is the boundary
twist parameter.

\begin{figure}
\centering
\includegraphics[angle=0,width=0.5\textwidth] {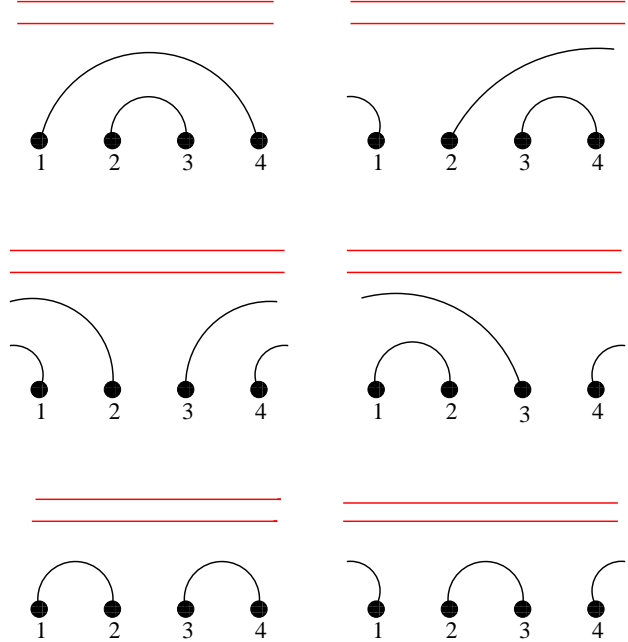}
\caption{(Color online)
 The six link pattern configurations for $L = 4$ sites on a cylinder
and two circles without sites (two noncontractible loops shown in red). The open arcs and
circles meet behind the cylinder.}
\label{prof1}
\end{figure}

\begin{figure}
	\centering
	\includegraphics[angle=0,width=0.5\textwidth] {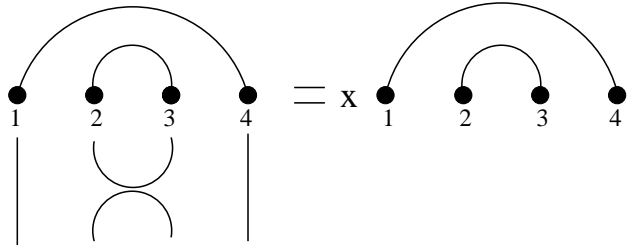}
\caption{
 The action of the $e_2$ generator acting on the bond between the sites
2 and 3 in one of the configurations appearing in Fig.~\ref{prof1}. The factor $x$
is due to the contractible loop.}
\label{prof2}
\end{figure}

\begin{figure}
	\centering
	\includegraphics[angle=0,width=0.5\textwidth] {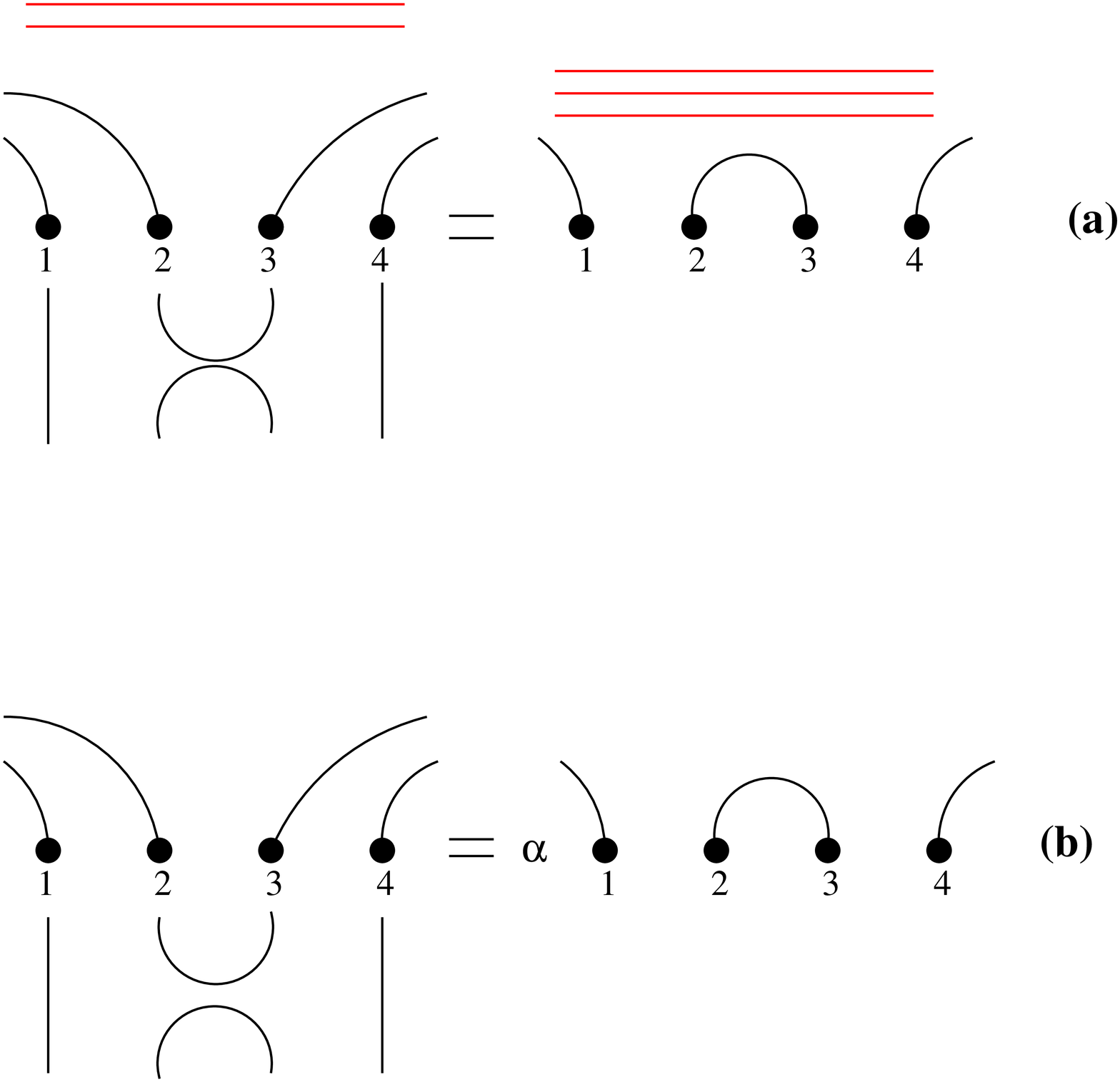}
\caption{(Color online)
(a) The action of the $e_2$ generator acting on the bond between the sites
2 and 3 which are the end points of an arc of the size of the system.
(b) The action of $e_2$ gives a factor $\alpha$ to the new configuration.}
\label{prof3}
\end{figure}

   In the second representation, the generators act in the vector space of
periodic link patterns. Each link pattern is one of the ${L} \choose {L/2}$
configurations of
nonintersecting arches joining $L$ sites on a circle. One can visualize the
circle on a cylinder. Besides the link patterns on the same
cylinder, one takes $m$ circles with no sites on them to
represent $m$ noncontractible loops. In Fig.~\ref{prof1} we show the six configurations
 for $L = 4$ and $m = 2$.
With few exceptions, the generators $e_i$ act on the configurations in the
standard way (as in the non-periodic Temperley-Lieb algebra) \cite{BBB}. In Fig.~\ref{prof2},
one sees the action of the generator $e_2$ on one of the configurations of Fig.~\ref{prof1}.
The factor $x$
appears due to a contractible loop. The exceptions occur if one considers
configurations having an arch of the size $L$ of the system and if the
generator acts
on the bond between the ends of the arch, see Fig.~\ref{prof3}(a). The action of
$e_2$ on the third configuration of Fig.~\ref{prof1} produces a new circle and one gets a
configuration with $m = 3$. Instead of considering configurations with various numbers of
noncontractible loops, we are going to consider configurations with no
noncontractible loops but, as a result of the action of $e_2$, we
multiply by fugacity $\alpha$ instead of adding a noncontractible loop; see Fig.~\ref{prof3}(b).
With this rule, one obtains a representation of the PTL with the quotient \rf{2.2}, and relates $\alpha$ to
the fugacity of noncontractible loops.

 We consider the Hamiltonian
\begin{equation} \label{2.6}
H = - \sum_{i = 1}^L e_i .
\end{equation}
Using the representation in terms of link paths of the PTL,
this Hamiltonian is equal up to a factor to the Hamiltonian $H'$ obtained from
the transfer matrix of the $O(n)$ models \cite{PRVill}. This factor is equal to the sound velocity
$v_s = \frac{\pi}{\gamma}\sin(\gamma)$:
\begin{equation} \label{2.7}
H' = H/v_s.
\end{equation}

In the spin representation of the PTL, using a similarity transformation, the
Hamiltonian $H'$  can be written as
\begin{eqnarray} \label{2.8}
H' = -\frac{1}{v_s}
\left[\sum_{i=1}^{L-1}\left(\sigma_i^+\sigma_{i+1}^- +\sigma_i^-\sigma_{i+1}^+
-\frac{\cos(\gamma)}{2}\, \sigma_i^z\sigma_{i+1}^z\right)\right. \nonumber \\
\left. +(\sigma_L^+\sigma_1^- e^{i\phi} + \sigma_L^-\sigma_1^+e^{-i\phi})
-\frac{\cos(\gamma)}{2}\, \sigma_L^z\sigma_1^z)+ \frac{L\cos(\gamma)}{2}\right].
\end{eqnarray}
This Hamiltonian which is the $XXZ$ quantum chain with a twist $\phi$ is
integrable and its ground-state and energy spectrum is known \cite{Alcaraz}. In particular, the
ground-state energy is
\begin{equation} \label{2.9}
E'(\phi,\gamma,L) = e_{\infty}'L +\left(\frac{\phi^2}{4(\pi-\gamma)} - \frac{\pi}
{6}\right) \frac{1}{L} +o(1/L),
\end{equation}
where $e_{\infty}'$ is the bulk energy density. Notice that the choice $\alpha=2$ ($\phi =0$)
and $\gamma=\pi/2$ used in Ref. \cite{PRVill} corresponds to the $XX$ model.
  From \rf{2.9} one can get two quantities of interest. Firstly, taking into
account that the density of the spin current is
\begin{equation} \label{2.10}
J_i={\rm i}\,(\sigma_i^+\sigma_{i+1}^- - \sigma_i^-\sigma_{i+1}^+),
\end{equation}
its average value at large values of $L$ is, see Eqs. \rf{2.4} and \rf{2.6}),
\begin{equation} \label{2.11}
J^z = -\frac{\partial E'} {\partial \phi} = - \frac{\phi}{2(\pi-\gamma)}
\frac{1}{L}.
\end{equation}
  Since the coefficient in Eq. \rf{2.11} is dimensionless, we expect it to be universal. A
second quantity of interest is the density of noncontractible loops:
\begin{equation} \label{2.12}
\rho_L(\gamma,\alpha) = -\alpha\frac{\partial E'}{\partial \alpha}=
-2\cot(\phi/2)J^z = \frac{\phi\cot(\phi/2)}{\pi-\gamma}\frac{1}{L}.
\end{equation}
  Notice that the density of noncontractible loops is proportional to the
current density (a physical explanation of this observation is still missing).
We observe also that the dependence on $n$ in $O(n)$ has a very simple form.

There is a simple way to check if the probability distribution of
noncontractible loops is Gaussian or not. Using Eqs. (\ref{2.5}) and (\ref{2.9}) one can
can compute all the moments $M_n$ of the probability distribution. They are
all of order $L^{-1}$. One can check that, for example, the identity
\begin{equation}\label{moments}
M_3 = 2M_1M_2 - M_1^3
\end{equation}
valid for a Gauss distribution, is not satisfied.

  Using the relation \rf{2.5}, one  obtains for $x = 0$ ($\gamma = \pi/2$,
no contractible loops)
\begin{equation} \label{2.13a}
\rho_L(\alpha) = \frac{4\alpha\arccos(\alpha/2)}
{\pi L\sqrt{4-\alpha^2}} \mbox{ for } \alpha \leq 2,
\end{equation}
and
\begin{equation} \label{2.13b}
\rho_L(\alpha) = \frac{4\alpha\mathrm{Arch}(\alpha/2)}
{\pi L\sqrt{\alpha^{2}-4}} \mbox{ for } \alpha > 2.
\end{equation}
Expressions \rf{2.13a} \rf{2.13b} are going to be compared to those
obtained from spanning webs model presented in the next sections.

For the same case ($x = 0$) only, one has a simple expression for the
ground state energy valid for any value of $L$:
\begin{equation}
E'(\phi, \pi/2, L) = - \cos(\phi/L)/\sin(\pi/L).
\end{equation}
Using this relation, one can compute the density of noncontractible loops
for any fugacity and any size of the system $L$.

\section{The spanning webs model on the rotated square lattice}

\begin{figure}[!ht]
\includegraphics[width=80mm]{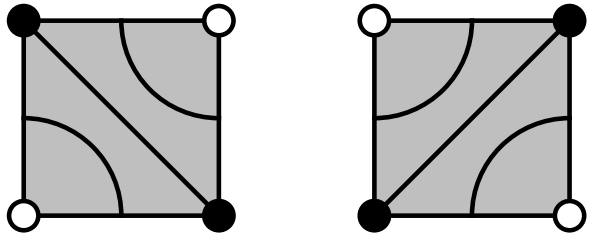}
\caption{\label{fig-bijection} The bijection between bonds and elementary cells.}
\end{figure}

\begin{figure}[!ht]
\includegraphics[width=80mm]{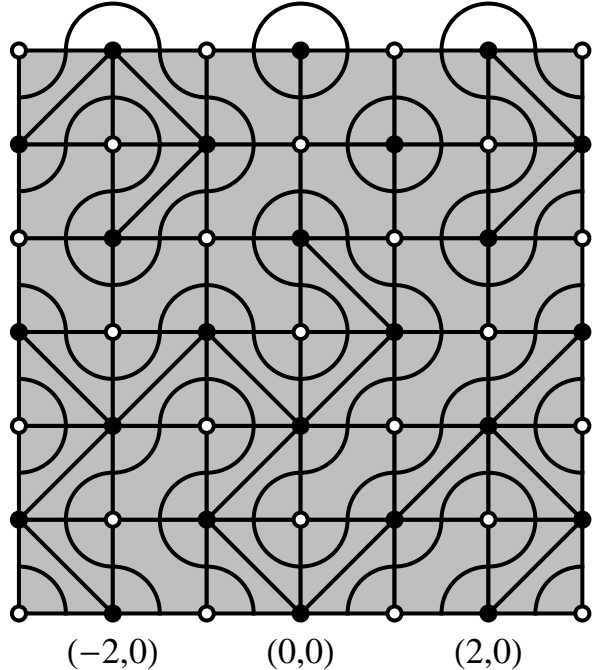}
\caption{\label{fig-loop-bonds} The bond configuration corresponding to the loop configuration from Fig.~\ref{fig-loops}.}
\end{figure}

Another representation of the $O(n)$ model relates loop configurations to clusters of bonds on sublattices of the original
lattice. The square lattice of sites with integer coordinates can be divided
into two sublattices, black and white. For sites of the black sublattice the sum of coordinates is
even, while for sites of the white one it is odd. The bijection between loop and bond configurations  is 
shown in Fig.~\ref{fig-bijection}.
The neighboring sites of each sublattice are connected by a bond if it does not intersect the borderlines of
the elementary cell. Each connected cluster of bonds in the bulk of the lattice is situated
inside a loop. Each bulk cluster on the black sublattice is surrounded by a connected
cluster of bonds on the white sublattice and vice versa.
The clusters of bonds corresponding to the loop configuration in Fig.~\ref{fig-loops} are
shown in Fig.~\ref{fig-loop-bonds}.

At $n=0$, the absence of bulk loops contractible to a point implies the absence of isolated clusters of bonds on the black or white sublattice. The allowed bond configurations are the so called spanning webs, i.e., the graphs containing all vertices of a sublattice, noncontractible loops arising from the periodic boundary conditions and spanning trees connected either to open boundaries or to the noncontractible loops. Clearly, due to the bijection between bonds and elementary cells, the presence of every noncontractible loop in the bulk of the cylinder is equivalent to the presence of a pair of noncontractible polymers in the dense polymer representation. If there are no noncontractible loops in the system, polymers can propagate from the top to the bottom of the cylinder. These polymers are treated as defect lines that separate clusters of bonds one from another.

The reformulation of the dense polymer model in terms of bond configurations leads us to the standard problem of spanning graphs on the square lattice. The cycle-free spanning graphs are called spanning trees; the graphs containing a number of cycles are called spanning webs. The enumeration of spanning trees is traced back to the classical Kirchhoff theorem \cite{Kirch,Pr85}. The spanning web model appears in statistical mechanics as the Temperley representation \cite{Temp} of the dimer model
solved by Kasteleyn \cite{Kastel} and by Temperley and Fisher \cite{Fisher}. A particular case we consider here is the spanning web model on a cylinder with noncontractible cycles supplied by fugacity $\xi$. This decoration needs a generalization
of the Kirchhoff theorem. A similar model, considered as the (1,2) logarithmic minimal model, has been solved in \cite{Tip}, where the fugacity of noncontractible loops was first introduced.
A basic accent of the present work is the calculation of the density of noncontractible cycles in a finite geometry.

\begin{figure}[!ht]
\includegraphics[width=100mm]{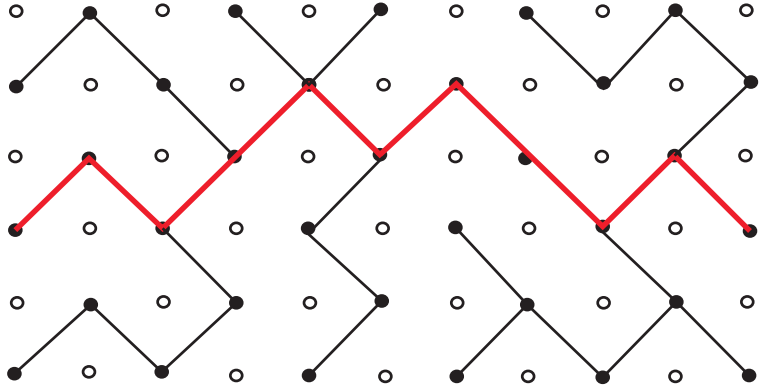}
\caption{(Color online) Spanning web with a single non-contractible loop (shown in red). }
\label{superlattice}
\end{figure}

We consider an oriented  labeled graph $\G =(V,E)$ with vertex set $V$ and
set of bonds $E$. Vertices are the sites of a
finite square lattice rotated by $\pi/4$ and wrapped on a cylinder.
The graph $\G =(V,E)$ can be  considered as a sublattice of square superlattice  $\L$ with
standard orientation, containing $N$ rows
and $L$ columns of cells. Vertices of the superlattice are shown in Fig.~\ref{superlattice}
as open and filled circles.
The vertex set $V$ is the sublattice of filled circles.
For convenience, in the remainder we call $N$ the ``height'' and $L$ the ``perimeter''
of the cylinder. Let the cells ${\bf r}_{ij}=(i,j)$ of $\L$ be
labeled by the integer coordinates $i=1,\dots, L$ and $j=1,\dots, N$,
so that the row $\{{\bf r}_{i1}=(i,1): i=1,\dots, L\}$ is the bottom
boundary of the cylinder and the row $\{{\bf r}_{iN}=(i,N): i=1,\dots, L\}$
is its top boundary. For the sake of convenience, both $L$
and $N$ are chosen to be even. The vertex
set $V$ of the rotated square lattice $\G$ then consists of the vertices of
the sublattice of $\L$ with, say, an even sum of the horizontal and vertical
coordinates, i.e., $V = \{{\bf r}_{ij} =(i,j): i+j=\mathrm{even}\}$.
Explicitly, we have
\begin{equation}\label{V}
V = \bigcup_{i=1}^{L/2}\bigcup_{j=1}^{N/2}\{(2i-1,2j-1),(2i,2j)\}.
\end{equation}
The edges in $E$ we take oriented from a site $(i,j)\in V$
to its nearest neighbors on the right-hand side, $(i+1,j+1)$ and
$(i+1,j-1)$. We call this direction ``positive'',
and the opposite one, from a site $(i,j)\in V$ to its nearest neighbors on the left-hand side,
$(i-1,j+1)$ and $(i-1,j-1)$, we call ``negative''.

We find it convenient to analyze the construction of spanning web configurations on
the above oriented graph by using the arrow representation; see, e.g.,
\cite{Pr85}. Accordingly, to each vertex ${\bf r} \in V$ we attach an
arrow directed along one of the bonds $({\bf r},{\bf r'})$ incident to it. Each
arrow defines a directed bond $({\bf r}\ra {\bf r'})$ and each configuration of
arrows $\A$ on $\G$ defines a spanning directed graph (digraph)
$\G_{sd}(\A)$  with set of bonds $E_{sd}(\A)=\{({\bf r}\ra {\bf r'}):
{\bf r}, {\bf r'}\in V\}$ depending on $\A$.

A cycle of length $k$ is a sequence of directed bonds
$({\bf r_1},{\bf r_2})$,$({\bf r_2},{\bf r_3})$, $({\bf r_3},{\bf r_4}), \dots,
({\bf r_k},{\bf r_1})$, where all ${\bf r_j}$,
$1 \leq j \leq k$ are distinct. If both $({\bf r}\ra {\bf r'})$ and
$({\bf r'}\ra {\bf r})$
belong to the same spanning web we say that it contains a cycle of length
2. Our aim is to study sets of spanning digraphs with no other
cycles than those which wrap the cylinder. The relevant
configurations will be enumerated with the aid of a generating
function defined as the determinant of an appropriately constructed
weight matrix:
\begin{equation}
Z_{L,N}(\omega) = \det D_{L,N}(\omega).
\label{ZD}
\end{equation}
In this respect, the derivation of the above relationship, see the
Appendix, is merely a generalization of the matrix Kirchhoff
theorem \cite{Kirch}.

The elements $\left(D_{L,N}(\omega)\right)_{\alpha,\beta}$, $\alpha,\beta =1,\dots, n=LN/2$ of the matrix $D_{L,N}$, 
associated with the graph $\G$, are explicitly given as
\begin{equation}
\left(D_{L,N}(\omega)\right)_{\alpha,\beta} = \left\{\begin{array}{rll}
z_\alpha,& \mathrm{if}& \; \alpha=\beta, \\
-b,&  \mathrm{if}& \; {\bf r}_\beta \; \mathrm{is} \;\mathrm{a} \; \mathrm{right} \; \mathrm{neighbor}
\; \mathrm{of} \; {\bf r}_\alpha,\\
- b^{-1},&  \mathrm{if}& \; {\bf r}_\beta \; \mathrm{is} \;\mathrm{a} \;  \mathrm{left}\; \mathrm{neighbor}
\; \mathrm{of} \; {\bf r}_\alpha,\\
0,& &\mathrm{otherwise}.
\end{array}      \right. \label{elementsD}
\end{equation}
Here $z_\alpha$ is the order of vertex ${\bf r}_\alpha$ in $\G$, $b= \omega^{1/L} e^{-i\pi/L}$, and the condition 
``${\bf r}_\beta$ is a right (left) neighbor of ${\bf r}_\alpha$'' means that if ${\bf r}_\alpha =(i,j)$,
then ${\bf r}_\beta =(i+1,j\pm 1)$ [${\bf r}_\beta =(i-1,j\pm 1)$].
Note that all closed paths which do not wrap the cylinder contain an equal
number of edges with either orientations, hence their
weight in $\det D$ remains the same as in $\det\Delta$.
Therefore, all the configurations that contain such closed paths
(contractible cycles) cancel out in the expansion of $\det
D$. On the other hand, cycles generated by off-diagonal
elements that wrap the cylinder change their sign, because they
contain edges oriented in one direction exceeding by
$L$ the number of edges in the opposite direction. This amounts to
the total factor of $b^L=-\omega$ or $b^{-L}=-\omega^{-1}$
depending on the orientation. Therefore, each noncontractible
cycle with a given orientation is counted twice, however, with
different weight --- once it enters into the determinant expansion
with the unit weight, being generated by diagonal elements of the
matrix $D$, and second time it enters with a factor
$\omega$ or $\omega^{-1}$ (depending on the orientation) as
generated by off-diagonal elements of that matrix. Thus, the total
number of noncontractible cycles, irrespective of their origin
and orientation, is given by the coefficient in front of the
corresponding power of $\omega + \omega^{-1}+2 \equiv \xi$ in the
series expansion of the partition function. In
general, besides the noncontractible cycles, the average number of
which is controlled by fugacity $\xi$, the spanning digraph
contains tree subgraphs connected to the cycles. All branches of the
trees can be generated only by the diagonal elements of
$D$ and, hence, carry unit weight.

\section{The partition function on a torus}

To calculate the generating function $Z_{L,N}(\omega)$ in expression (\ref{ZD}), we
make some initial transformations that allow us to easily  diagonalize
the matrix $D_{L,N}(\omega)$. First of all, the decomposition
(\ref{V}) of the vertex set $V$ of the rotated square lattice $\G$
suggests its rearrangement by
combining all pairs of nearest neighbors $(2i-1,2j-1)$ and
$(2i,2j)$, $i=1,2,\dots, L'=L/2$, $j=1,2,\dots, N'=N/2$ into
two-site unit cells. Thus we obtain a square $L'\times N'$ array of
$LN/4$ unit cells with the connectivity of a triangular lattice.
Under neglect of the boundary effects at the top and bottom of
the cylinder, we describe the weighted connectivity of the sites
in a unit cell with their neighbors in $\G$, taking into account the bond
orientation, by introducing the following $2\times 2$ matrices:
$$
a(0,0) = \left( \begin{array}{cc}
0&b^{-1}\\ b&0 \end{array}\right), \quad
a(1,0) = \left( \begin{array}{cc}
0&0\\b^{-1}&0 \end{array}\right), \quad
a(0,1) = \left( \begin{array}{cc}
0&0\\b&0 \end{array}\right), \quad
a(1,1) = \left( \begin{array}{cc}
0&0\\b^{-1}&0 \end{array}\right),
$$
$$
a(-1,0) = \left( \begin{array}{cc}
0&b\\0&0 \end{array}\right), \quad
a(0,-1) = \left( \begin{array}{cc}
0&b^{-1}\\0&0 \end{array}\right), \quad
a(-1,-1) = \left( \begin{array}{cc}
0&b\\ 0&0 \end{array}\right). \label{a}
$$

Now the matrix $D_{L,N}(\omega)$, see Eq. (\ref{elementsD}), can
be written as
\begin{eqnarray}
&&[4 I_2 - a(0,0)]\otimes I_{L'}\otimes I_{N'} - a(1,0)\otimes R_{L'}\otimes I_{N'}
- a(-1,0)\otimes R_{L'}^T\otimes I_{N'}- a(0,1)\otimes I_{L'}\otimes R_{N'} \nonumber
\\&& - a(0,-1)\otimes I_{L'}\otimes R_{N'}^T - a(1,1)\otimes R_{L'}\otimes R_{N'}
- a(-1,-1)\otimes R_{L'}^T\otimes R_{N'}^T.
\end{eqnarray}
Here $R_M$ is the $M\times M$ matrix
\begin{equation}
R_M = \left( \begin{array}{ccccc}
0&1&0&\dots&0\\ 0&0&1&\dots&0 \\
\dots&\dots&\dots&\dots&\dots \\ 0&0&\dots&0&1\\ 1&0&\dots&0&0
 \end{array}\right), \label{R}
\end{equation}
and $R_M^T$ is the matrix transposed of $R_M$. Now we note that
both $R_M$ and $R_M^T$ are diagonalized by the similarity transformation
$$S_M^{-1}R_M S_M = \mathrm{diag}\{\mathrm{e}^{\mathrm{i}2\pi m/M}, m=1,2,\dots,M\},$$
where the $S_M$ is the matrix with elements
\begin{equation}
\left(S_M\right)_{n,m} = M^{-1/2}\mathrm{e}^{\mathrm{i}2\pi m n/M},\quad m,n=1,2,\dots,M.
\end{equation}
Since $R_M R_M^T =I_M$, we have
$$S_M^{-1}R_M^T S_M = \mathrm{diag}\{\mathrm{e}^{-\mathrm{i}2\pi m/M}, m=1,2,\dots,M\}.$$
Therefore, with the similarity transformation generated by the matrix $I_2\otimes S_{L'}\otimes
S_{N'}$ we can diagonalize the matrix $D_{L,N}(\omega)$ in the $L'=L/2$ and $N'=N/2$-
dimensional subspaces. Then for the determinant we readily obtain
\begin{equation}\label{detD}
\det D_{L,N}(\omega)= \prod_{m=1}^{L'}\prod_{n=1}^{N'}\det Q(2\pi m/L', 2\pi n/N'),
\end{equation}
where $Q(\theta_1,\theta_2)$ is the $2\times 2$ matrix,
\begin{eqnarray}
&& Q(\theta_1,\theta_2)= 4 I_2 - a(0,0) - a(1,0)\mathrm{e}^{\mathrm{i}\theta_1}
- a(-1,0)\mathrm{e}^{-\mathrm{i}\theta_1}- a(0,1)\mathrm{e}^{\mathrm{i}\theta_2} \nonumber
\\&& - a(0,-1)\mathrm{e}^{-\mathrm{i}\theta_2} - a(1,1)\mathrm{e}^{\mathrm{i}\theta_1 +
\mathrm{i}\theta_2}
- a(-1,-1)\mathrm{e}^{-\mathrm{i}\theta_1 -\mathrm{i}\theta_2}.
\end{eqnarray}
It is convenient to cast its determinant in the form
\begin{equation}
\det Q(\theta_1,\theta_2)=4 \cos^2(\theta_2/2)\left[\frac{4}{\cos^2(\theta_2/2)}-2 -
2\cos(\theta_1 +\delta)\right],
\end{equation}
where $\delta = 2\mathrm{i}\ln b = (\mathrm{i}\ln \omega +\pi)/L'$.
Thus, from Eq. (\ref{detD}) we obtain the $\omega$-dependent part of the
partition function
\begin{equation}\label{Z}
Z_{L',N'}(\omega)= \prod_{m=1}^{L'}\prod_{n=1}^{N'}\left[\frac{4}{\cos^2(\pi n/N')}-2 -
2\cos\left(2\pi m/L' +\delta \right)\right].
\end{equation}

Next we set
\begin{equation}\label{A2}
\frac{4}{\cos^2(\pi n/N')}-2 = A^2(\pi n/N') + A^{-2}(\pi n/N'),
\end{equation}
where
\begin{equation}\label{A}
A^2(\pi n/N')= \frac{\left(1 +|\sin(\pi n/N')|\right)^2}{\cos^2(\pi n/N')}.
\end{equation}
Now, making use of the identity
\begin{equation}
\prod_{m=1}^{L'}\left[A^2+A^{-2}-
2\cos\left(2\pi m/L'+\delta \right)\right]=
A^{2L'}+ A^{-2L'}-2 \cos(L'\delta),
\label{iden}
\end{equation}
with $\delta = (\mathrm{i}\ln \omega +\pi)/L'$, we perform exactly
the product over $m$:
\begin{equation}\label{Z1}
Z_{L',N'}(\omega)= \prod_{n=1}^{N'}\left[
\frac{\left(1 +\sin(\pi n/N')\right)^L}{\cos^L(\pi n/N')} +
\frac{\left(1 - \sin(\pi n/N')\right)^L}{\cos^L(\pi n/N')}
+\omega + \omega^{-1}\right].
\end{equation}

Strictly speaking, this expression is valid for spanning webs on
a torus, since we have not considered boundary conditions at the edges of the cylinder properly.
For the sake of completeness, in the next section we present the exact expressions for the partition
function under closed and open boundary conditions at the top and bottom of a finite-size cylinder.

\section{Cylindrical boundary conditions}

\begin{figure}[!ht]
\begin{centering}
\includegraphics[width=0.9\textwidth]{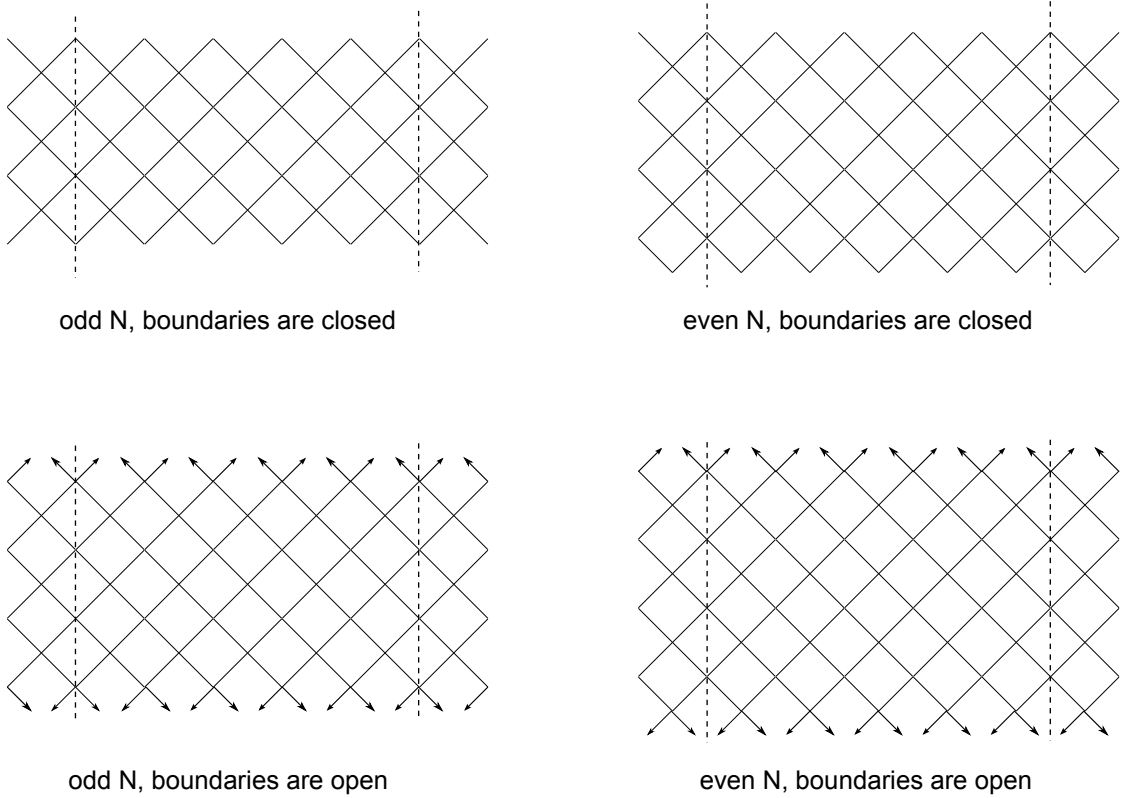}
\par\end{centering}
\caption{Rotated square lattices with different boundary conditions on the cylinder. The two dashed vertical
lines in each of the four panels represent a single line due to periodicity
in the horizontal direction. For example, the top-left lattice has $L=10$ and $N=7$.}
\label{fig:graphs}
\end{figure}

In the preceding section we computed the partition function of the spanning webs without contractible
loops on the torus. Here we consider the cases when the boundaries at the top and bottom of cylinder
are closed or open; see Fig.(\ref{fig:graphs}). It can be seen from the figure
that there is a difference between the cases of odd and even $N$: in the
latter case, the lower edge is shifted with respect to the top one. In our
considerations $L=2L'$ is even, whereas $N$ is of any parity.

In the case of both boundaries open, the diagonal elements
of the matrix (\ref{elementsD})are  $z_{\alpha}=4$, because every vertex has
four outgoing edges. The partition function is, compare to (\ref{Z}),
\begin{equation}
\label{Zoo}
Z_{L,N}^{\text{op,op}}=\begin{cases}
2^{L'}(N+1)^{L'}\,\prod\limits _{n=1}^{(N-1)/2}\prod\limits _{m=0}^{L'-1}\left(\dfrac{4}{\sin^{2}\frac{\pi m}{N+1}}-2- 2\cos(2\pi m/L')\right), & \text{odd }N\\
\prod\limits _{n=1}^{N/2}\prod\limits _{m=0}^{L'-1}\left(\dfrac{4}{\sin^{2}\frac{\pi(n-1/2)}{N+1}}-2-
2\cos(2\pi m/L'+\delta)\right), & \text{even }N.
\end{cases}
\end{equation}

The partition functions in the case of all the other uniform boundary conditions can be expressed in
terms of $Z_{L,N}^{\text{op,op}}$ as follows:

In the case of both closed boundary conditions,
\begin{equation}
Z_{L,N}^{\text{cl,cl}}=\xi Z_{L,N-2}^{\text{op,op}},\label{Zcc}
\end{equation}
in the case mixed open-closed boundary conditions,
\begin{equation}
Z_{L,N}^{\text{op,cl}}=
2^{-L'}\frac{Z_{L,2N-1}^{\text{op,op}}}{Z_{L,N-1}^{\text{op,op}}},
\end{equation}
and for the torus
\begin{equation}
Z_{L,N}^{\text{torus}}=4^{L'}\xi\left(Z_{L,N/2-1}^{\text{op,op}}\right)^{2}.
\end{equation}

From these expressions, it can be easily seen that the density of noncontractible loops, see the next section, is equal for 
different boundary conditions in the limit $N\rightarrow \infty$.

\section{Density of non-contractible loops}

The average density (per unit height of the cylinder) of the noncontractible cycles as
a function of the fugacity $\xi$ is defined as
\begin{equation}\label{density}
\rho_{L,N}(\xi) = \frac{1}{N}\xi \frac{\partial}{\partial \xi}\ln
\det D_{L,N}(\omega(\xi)).
\end{equation}

Since we are interested in the case of the density on an infinitely
long cylinder, we expect the boundary corrections to vanish in
the limit $N'\rightarrow \infty$. To take this limit, we factor
out another $\omega$-independent term from $Z_{L',N'}(\omega)$,
\begin{eqnarray}\label{Z2}
Z_{L',N'}(\omega)&=& \left[\prod_{n=1}^{N'}
\frac{[1 +\sin(\pi n/N')]^L}{\cos^L(\pi n/N')}\right]\nonumber \\
&\times& \prod_{n=1}^{N'}\left[1 + (\omega + \omega^{-1})\frac{\cos^L(\pi n/N')}
{[1 + \sin(\pi n/N')]^L}+ \frac{\cos^{2L}(\pi n/N')}
{[1+ \sin(\pi n/N')]^{2L}} \right],
\end{eqnarray}
and make use of the density definition (\ref{density}):
\begin{equation}\label{dens1}
\rho_{L,N}(\xi) = \frac{\xi \partial}{\partial \xi}\frac{1}{N}
\sum_{n=1}^{N'}\ln \left[1 + (\xi-2)\frac{\cos^L(\pi n/N')}
{[1 + \sin(\pi n/N')]^L}+ \frac{\cos^{2L}(\pi n/N')}
{[1+ \sin(\pi n/N')]^{2L}} \right].
\end{equation}
Hence, in the limit $N'\rightarrow \infty$
\begin{eqnarray}\label{dens2}
\rho_{L,\infty}(\xi) &=& \xi \frac{\partial}{\partial \xi}\frac{1}{2\pi}
\int_{0}^{\pi}\mathrm{d}\, \phi\ln \left[1 + (\xi-2)\frac{\cos^L(\phi)}
{[1 + \sin(\phi)]^L}+ \frac{\cos^{2L}(\phi)}
{[1+ \sin(\phi)]^{2L}} \right]\nonumber \\
&=&\frac{\xi}{2\pi}\int_{0}^{\pi}\frac{g^L(\phi)\; \mathrm{d}\, \phi}{1
 + (\xi-2)g^L(\phi) + g^{2L}(\phi)},
\end{eqnarray}
where ($L = 2L'$ is even)
\begin{equation}\label{g}
g(\phi \in [0,\pi]) =  \frac{|\cos(\phi)|}{[1+ \sin(\phi)]} \in [0,1], \quad
\max_{\phi \in [0,\pi]}g(\phi)= g(0)=g(\pi)=1.
\end{equation}

Therefore, when $L\gg 1$, the essential contribution in $\rho_{L,\infty}(\xi)$
comes from the integration over the two small intervals $0 \leq \phi \le \epsilon$
 and $\pi - \epsilon \le \phi \leq \pi$ with $\epsilon \ll 1$. To leading-order
in $L\gg 1$, it suffices to take the linear term in the expansion
\begin{equation}\label{gexp}
\ln g(\phi) = - \phi  + O(\phi^2), \quad
\phi \rightarrow 0^+, \quad \mathrm{or} \quad \phi \rightarrow \pi,
\end{equation}
which yields
\begin{eqnarray}\label{dens3}
\rho_{L,\infty}(\xi) &\simeq& \frac{\xi}{\pi}\int_{0}^{\epsilon}\frac{\mathrm{e}^{-L\phi}\;
\mathrm{d}\, \phi}{1 + (\xi-2)\mathrm{e}^{-L \phi} + \mathrm{e}^{-2L\phi}},\nonumber \\
&=&\frac{\xi}{\pi L}\int_{\mathrm{e}^{-L \epsilon}}^{1}\frac{\mathrm{d}\,y}{1 + (\xi-2)y + y^{2}}.
\end{eqnarray}
Extending the lower limit in the latter integral to $y=0$, we finally obtain the general
expression
\begin{equation}\label{final}
\rho_{L,\infty}(\xi) = \left\{\begin{array}{lll}
\frac{\xi}{\pi L \sqrt{\xi(\xi -4)}}\ln \frac{\xi + \sqrt{\xi(\xi -4)}}{\xi - \sqrt{\xi(\xi -4)}},&
\mathrm{if}& \; (\xi-4)>0, \\
\frac{2}{\pi L},&  \mathrm{if}& \; (\xi-4)=0,\\
\frac{2\xi}{\pi L\sqrt{\xi(4- \xi)}}\left[\tan^{-1} \frac{\xi}{\sqrt{\xi(4 -\xi)}}
- \tan^{-1} \frac{\xi -2}{\sqrt{\xi(4 -\xi)}}\right],&
\mathrm{if}& \; (\xi-4)< 0.\end{array}      \right.
\end{equation}

We are going to compare the obtained result with Eq. (\ref{2.13a}) in the interval of the loop fugacity $0 \leq \alpha \leq 2$
corresponding to real values of the twist parameter $\phi$.
Toward that end, we transform the bottom expression in Eq.(\ref{final}),
\begin{equation}\label{fintrans}
\frac{2\xi}{\pi L\sqrt{\xi(4- \xi)}}\left[\tan^{-1} \frac{\xi}{\sqrt{\xi(4 -\xi)}}
- \tan^{-1} \frac{\xi -2}{\sqrt{\xi(4 -\xi)}}\right]=\frac{2\xi}{\pi L \sqrt{\xi(4-\xi)}}\arccos\left(\frac{\sqrt \xi}{2}\right)
\end{equation}
and remember that each cycle in the spanning web model corresponds to a pair of non-contractible loops with fugacity
$\xi=\alpha^2$. Then, the density of non-contractible loops in the dense polymer model is
\begin{equation}\label{densloops}
\rho_L(\alpha)=\frac{4\alpha \arccos(\alpha/2)}{\pi L \sqrt{(4-\alpha^2)}}
\end{equation}
in full agreement with (\ref{2.13a}).

For $\xi > 4$ and $\alpha > 2$, we use the formula
\begin{equation}\label{formula}
\ln [(1+x)/(1-x)] =2\mathrm{Arth}(x)
\end{equation}
and get
\begin{equation}\label{densloops2}
\rho_L(\alpha)=\frac{4\alpha \mathrm{Arch}(\alpha/2)}{\pi L \sqrt{(\alpha^2-4)}}.
\end{equation}
We see that the formula for $\alpha > 2$ corresponds to the quantum chain result (\ref{2.13b}) with the complex twist
\begin{equation}\label{comtwist}
\phi = i\Omega,\ \ \ \ \alpha = 2 \cosh(\Omega/2).
\end{equation}

The crucial check of Eq.(\ref{densloops}) is the value of density $\rho_L(\alpha)$
for $\alpha=\sqrt 2$, when the fugacity of
noncontractible cycles in the spanning web model is $\xi = 2$. In this case, the
noncontractible cycles enter into the
partition function as two nonweighted sequences of bonds oriented clockwise and anticlockwise. Due to symmetry of the $O(n)$
model, the spread of each sequence in the horizontal direction (that is, $L$ by the definition of loops) coincides 
in average with that in the direction of the cylinder axis. Each two loops on the rotated lattice (Fig.\ref{superlattice}) are separated by a loop on the dual lattice. Thus, the spanning web configuration in the vertical direction is a sequence of sandwiches of
loops and dual loops of average thickness $2L$. Then, the density of web cycles is $1/(2L)$ and the density of
noncontractible loops is $\rho_L(\sqrt 2) = 1/L$.

We note that Morin-Duchesne {\it et al.} \cite{MD} have obtained the inversion relation for any $\alpha$ which should allow
one to obtain our results in a different way.

\section*{Acknowledgments}
We thank P. Pearce for helpful discussions. This work was supported by RFBR Grant No 12-01-00242a, the
Heisenberg-Landau program, the DFG grant No. RI 317/16-1 and the Brazilian agencies FAPESP and CNPq.

\section*{Appendix}

In this Appendix, we give a detailed derivation of the expression (\ref{ZD}) for the generating function
of all spanning digraphs on $\G$ which have no contractible cycles.
We begin with an examination of the determinant expansion of the
usual Laplace matrix $\Delta$ for the graph $\G$.
Let the vertices ${\bf r}\in V$ be labeled in arbitrary order from 1
to $n=|V|=LN/2$. Then $\Delta$ has the following elements
($\alpha,\beta \in \{1, \ldots, n\}$)
\begin{equation}
\label{laplacian}
\Delta_{\alpha,\beta} = \left\{ \begin{array}{rll}
z_\alpha,& \mathrm{if}& \; \alpha =\beta, \\
-1,&  \mathrm{if}& \; \alpha \; \mathrm{and}\; \beta \; \mathrm{are}\,
\mathrm{adjacent},\\
0,& &\mathrm{otherwise}.
\end{array}      \right.
\end{equation}
where $z_\alpha$ is the order of vertex ${\bf r}_\alpha$ in the rotated square
lattice $\G$. Since the matrix $\Delta$
has a zero eigenvalue, its determinant vanishes. On the other
hand, the Leibnitz formula expresses the determinant of $\Delta$ as
a sum over all permutations $\sigma$ of the set $\{1, 2,\dots ,
n\}$:
\begin{equation}
\det \Delta = \sum_{\sigma \in S_n} \; \mathrm{sgn} (\sigma) \:
\Delta_{1,\sigma(1)} \Delta_{2,\sigma(2)} \ldots
\Delta_{n,\sigma(n)} =0,
\label{Leibniz}
\end{equation}
where $S_n$ is the symmetric group and $\mathrm{sgn} (\sigma)=\pm 1$
is the signature of the permutation $\sigma$. The identity
permutation $\sigma = \sigma_{\mathrm{id}}$ in Eq. (\ref{Leibniz}) yields
the term $z_1 z_2 \cdots z_n$ equal to the number of all possible arrow configurations
on $\G$.

In general, each permutation $\sigma \in S_n$ can be factored into a product
(composition) of disjoint cyclic permutations, say $\sigma = c_1
\circ c_2  \cdots \circ c_k$. This representation partitions the
set of vertices $V$ into non-empty disjoint subsets --- the orbits
$\O_i$ of the corresponding cycles $c_i$, $i=1,\dots ,k$. More
precisely, if $\O_i=\{v_{i,1},v_{i,2}, \dots v_{i,l_i}\}
\subset V$ is the orbit of $c_i$, then $\cup_{i=1}^k\O_i
=V$ and $\sum_{i=1}^k l_i =n$, where $l_i$ is the
cardinality of the orbit $\O_i$, or equivalently, the length
of the cycle $c_i$. The orbits consisting of just one
element, if any, constitute the set $S_{\rm fp}(\sigma)$ of fixed
points of the permutation: $S_{\rm fp}(\sigma) = \{v=\sigma (v), v
\in V\}$. In the case of the identity permutation $\sigma_{\mathrm{id}} \in S_n$,
all orbits consist of exactly one element,
$\O_i(\sigma_{\mathrm{id}})=\{v_i\}\subset V$, $i=1, \dots ,n$, and $S_{fp}
(\sigma_{\mathrm{id}})= V$.
A cycle $c_i$ of length $|c_i|=l_i \geq 2$ will be called a
\textit{proper cycle}. A proper cycle of length 2 corresponds
to two oppositely directed edges that connect a pair of adjacent
vertices: $(v_{i,1}\ra v_{i,2})$, $(v_{i,2}\ra v_{i,1})$. Note that the vertices
of an orbit $\O_i$ of cardinality $l_i = |O_i(\sigma)| \geq 3$ are connected
by a closed path on $\G$ which can be traversed in two opposite
directions: if $c_i$ is the cycle defined by $v_{i,1} \ra \sigma(v_{i,1}) =
v_{i,2}, \ra \dots \ra \sigma(v_{i,l_i})= v_{i,1}$, then the reverse
cycle $c'_i$ can be represented as $v_{i,l_i} \ra
\sigma(v_{i,l_i}) = v_{i,l_i -1}, \ra \dots \ra \sigma(v_{i,1})=
v_{i,l_i}$.

Now we take into account that the proper cycles on $\G$ are of
even length only, hence the signature of every permutation in the
expansion of the determinant depends on the number of proper
cycles in its factorization, i.e., if $\sigma = c_1 \circ c_2
\cdots \circ c_p$, where $|c_i|\geq 2$, $i=1,\dots , p$, then
$\mathrm{sgn} (\sigma)= (-1)^p$. Thus, the terms in Eq.
(\ref{Leibniz}) can be rearranged according to the number $p$ of
disjoint proper cycles as follows:
\be
\prod_{i=1}^n z_i +  \sum_{p=1}^{[n/2]} (-1)^{p} \sum_{\sigma = c_1 \circ
\cdots \circ c_p} \; \prod_{i=1}^{p}
\Delta_{v_i,c_i(v_i)} \Delta_{c_i(v_i),c_i^2(v_i)}\cdots
\Delta_{c_i^{l_i -1}(v_i),v_i}\prod_{j\in S_{fp}(\sigma)}z_j.
\label{altern}
\ee
Here $c_i^k$ is the $k$-fold composition of the cyclic permutation
$c_i$ of even length $l_i$, $v_i \in \O_i(\sigma)$, so that
$c_i^{k-1}(v_i)\not= c_i^k (v_i)$ and $c_i^{l_i}(v_i) = v_i$.
Note that all nonvanishing off-diagonal elements are equal to $-1$.

The above expansion
reveals the following features: (i) As expected, all spanning digraphs on
$\G$ have at least one proper cycle; (ii) Each term with $S_{\rm fp}(\sigma)
\not= \emptyset$ represents a set of $\prod_{j\in S_{\rm fp}}z_j$ distinct
spanning digraphs, which have in common the specified cycles $c_1,\dots , c_p$,
and they differ in the oriented edges outgoing from the vertices
$j\in S_{\rm fp}(\sigma)$. These oriented edges may form cycles on
their own which do not enter into the list $c_1,\dots , c_p$; (iii) Since
the sets $\cup_{i=1}^p\O_i$ and $S_{\rm fp}(c_1,\dots , c_p)$ are disjoint,
the proper cycles formed by the oriented edges incident to the fixed points of
a given permutation $\sigma = c_1\circ c_2 \circ \cdots \circ c_p$ should
enter into the enlarged list of cycles $c_1, c_2, \dots, c_p,\dots, c_{p'}$,
$p'>p$, corresponding to the cycle decomposition of another permutation $\sigma'$.

\begin{figure}[!ht]
\includegraphics[width=80mm]{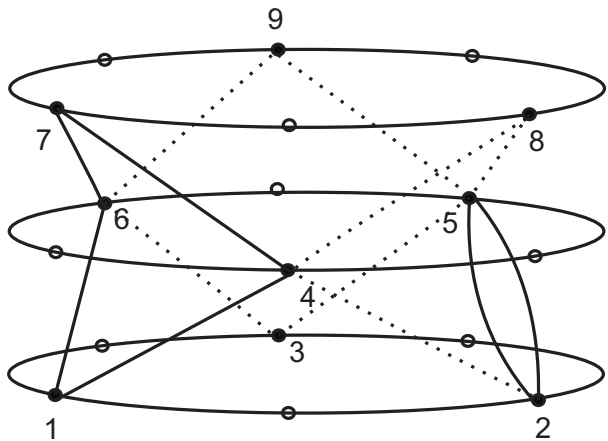}
\caption{Spanning digraph generated by a single term in the determinant expansion of the Laplacian matrix (see the text). }
\label{example}
\end{figure}

For example, consider the determinant of the Laplacian matrix of a cylinder of height 3
and perimeter 6 shown in Fig. \ref{example}. The set of vertices $V$ consists of filled circles
marked by $1,2,\dots,9$. The set of oriented edges is a collection of 24 inclined vectors of
type $1 \rightarrow 4,1 \rightarrow 6, 4 \rightarrow 1, 6 \rightarrow 1,\dots $. The
corresponding Leibnitz expansion contains the term
\begin{equation}
\label{LeibnExam1}
(-1)^2 (\Delta_{1,4}\Delta_{4,7}\Delta_{7,6}\Delta_{6,1}) %
       (\Delta_{2,5}\Delta_{5,2})\Delta_{3,3}\Delta_{8,8}\Delta_{9,9}
\end{equation}
which represents $z_{3}z_{8}z_{9}= 8$ spanning digraphs on $\G$ with 2 specified
cycles and all possible oriented bonds outgoing from the vertices 3,8 and 9.

As noticed first in Ref. \cite{Pr85}, the expansion (\ref{altern}) parallels in form the
\textit{inclusion-exclusion principle} in combinatorial mathematics. Indeed, let
$c_1, c_2, \dots, c_m$ be the list of all possible proper cycles on $\G$,
labeled in an arbitrary order. Define $A_i$, $i=1,2,\dots ,m$ as
the set of all spanning digraphs on $\G$ containing the particular cycle $c_i$.
Then, expansion (\ref{altern}) can be written in the form of the inclusion-exclusion
principle:
\begin{equation}
\left| \cup_{i=1}^m A_i \right| - \sum_{i=1}^m|A_i| + \sum_{1\leq i<j \leq m} |A_i \cap A_j|
 - \sum_{1\leq i<j<k \leq m} |A_i\cap  A_j \cap A_k|+ \cdots
 -(-1)^{m+1}|A_1 \cap \cdots \cap A_m|, \label{inex}
\end{equation}
which holds for any finite sets $A_1, A_2, \dots , A_m$,
where $|A|$ is the cardinality of the set $A$. This sum equals zero, because all spanning
digraphs on $\G$ have at least one proper cycle $c_i$, $i=1,2,\dots, m$. The first term of
the expansion originates
from the term  $\prod_{i=1}^n z_i$ in Eq. (\ref{altern}) and represents the set of all
possible cycles formed by oriented edges incident to every vertex of $\G$. To obtain the
number of spanning webs, one has to subtract all the digraphs having contractible cycles
and leave all those with noncontractible cycles wrapping the cylinder. To
keep the number of noncontractible cycles, we have to change the weights of the nondiagonal terms
in such a way that every cycle containing difference between the numbers of positive and
negative steps equal to $+L$ or $-L$, depending on the orientation,
enters the determinant expansion with the opposite sign and sums up with the corresponding
non-contractible cycle, generated by the arrows representing the diagonal
term $\prod_{i=1}^n z_i$. At that, all the
proper contractible cycles keep their sign in the expansion (\ref{altern}) in order to
cancel out. This readily follows from the fact that every contractible cycle contains
equal numbers of positive and negative steps. Thus, the matrix $D_{L,N}$ with elements
given by (ref{elementsD}) has all the necessary properties for the relationship
(\ref{ZD}) to define the proper generating function.


\begin{thebibliography}{99}


\bibitem{dense} H.W.J. Bl{\"o}te and B. Nienhuis, J. Phys. A {\bf 22}, no. 9, 1415 (1989).

\bibitem{Batch} M.T. Batchelor, J. de Gier and B. Nienhuis, J. Phys. A {\bf 34}, no.19, L265 (2001).

\bibitem{Mitra} S. Mitra, B. Nienhuis, J. de Gier and  M.T. Batchelor, J. Stat. Mech., P09010 (2004).

\bibitem{Pearce} P.A. Pearce and J. Rasmussen, J. Stat. Mech., P02015 (2007).

\bibitem{Saleur-1}H. Saleur, J. Phys. A {\bf 20}, 455 (1987).

\bibitem{Dupl}B. Duplantier, J. Phys. A {\bf 19}, L1009 (1986).

\bibitem{Saleur-2}H. Saleur, Nucl. Phys. B {\bf 382}, 486 (1992).

\bibitem{PRVill} P. A. Pearce, J. Rasmussen and S. Villani, J. Stat. Mech., P02010 (2010).

\bibitem{Izmail} N. Sh. Izmailian, V. B. Priezzhev, Ph. Ruelle, and C.-K. Hu, Phys. Rev. Lett. {\bf 95},
260602 (2005).

\bibitem{Levy} D. Levy, Phys. Rev. Lett. {\bf67}, 1971 (1991); Int. J. Mod. Phys. A {\bf 6}, 5127 (1991).
\bibitem{AAA}
A. M. Gainutdinov, J. L. Jacobsen, N. Read, H. Saleur and R. Vasseur,
J. Phys. A {\bf 46}, 494012 (2013) and references therein;
F. C. Alcaraz and V. Rittenberg, J. Stat. Mech., P09010 (2013);
F. C. Alcaraz, P. Pyatov and V. Rittenberg, J. Phys. A {\bf 47}, 462001 (2014).

\bibitem{BBB}
 P. Martin, {\it Potts models and related problems in statistical mechanics}, World Scientific (1990)

\bibitem{Alcaraz} F. C. Alcaraz, M. N. Barber, and M. T. Batchelor, Ann. Phys. (NY) {\bf 182}, 280 (1988).

\bibitem{Kirch} C. Kirchhoff, Ann. Phys. Chem. {\bf 148}, 497 (1847).

\bibitem{Temp} H. N. V. Temperley, in: London Math. Soc. Lecture Note Series, vol 13, Cambridge University Press,
Cambridge, 1974, p.202.

\bibitem{Kastel}P. W. Kasteleyn, Physica {\bf 27}, 1209 (1961).

\bibitem{Fisher} H. N. V. Temperley and M. E. Fisher, Philos. Mag. {\bf 6}, 1061 (1961).

\bibitem{Tip} J. G. Brankov, S. Yu. Grigorev, V. B. Priezzhev, I. Y. Tipunin, J. Stat. Mech., P11017 (2008).

\bibitem{Pr85} V. B. Priezzhev, Sov. Phys. Usp. {\bf 28}, 1125 (1985).

\bibitem{MD} A. Morin-Duchesne, P. A. Pearce, and J. Rasmussen, Nucl. Phys. B {\bf 874}, 312 (2013).

\end{thebibliography}
\end{document}